\documentclass[conference]{IEEEtran}
\usepackage[utf8]{inputenc}
\usepackage{graphicx}
\usepackage{xcolor}
\usepackage{url}
\usepackage{mdframed}
\usepackage{subfig}
\usepackage{tabularx}
\usepackage{multirow}
\usepackage{colortbl}
\usepackage{booktabs} 
\usepackage{color}
\usepackage{tabularx}
\usepackage{soul}

\begin{document}
\title{OpenReq Issue Link Map:  A Tool to Visualize Issue Links in Jira}

\author{
\IEEEauthorblockN{Clara Marie L\"{u}ders}
\IEEEauthorblockA{The Qt Company\\
Espoo, Finland\\
clara.luders@qt.io}
\and
\IEEEauthorblockN{Mikko Raatikainen}
\IEEEauthorblockA{University of Helsinki\\
Helsinki, Finland\\
mikko.raatikainen@helsinki.fi}
\and
\IEEEauthorblockN{Joaquim Motger}
\IEEEauthorblockA{Polytechnic University of Catalonia\\
Barcelona, Spain\\
jmotger@essi.upc.edu}
\and
\IEEEauthorblockN{Walid Maalej}
\IEEEauthorblockA{University of Hamburg\\
Hamburg, Germany\\
maalej@informatik.uni-hamburg.de}
}
\maketitle

\begin{abstract}
Managing software projects gets more and more complicated with an increasing project and product size.
To cope with this complexity, many organizations use issue tracking systems, where tasks, bugs, and requirements are stored as issues. Unfortunately, managing software projects might remain chaotic even when using issue trackers. Particularly for long lasting projects with a large number of issues and links between them, it is often hard to maintain an overview of the dependencies, especially when dozens of new issues get reported every day. We present a Jira plug-in that supports developers, project managers, and product owners in managing and overviewing issues and their dependencies. Our tool visualizes the issue links, helps to find missing or unknown links between issues, and detects inconsistencies.
\end{abstract}
\IEEEpeerreviewmaketitle

\section{Introduction}
Many software companies keep track of their work in issue tracking systems, such as Jira, Bugzilla, or Github. Stakeholders can describe and store requirements---called issues or tickets in these systems---in the form of epics, user stories, tasks, bugs, feature requests or other defined types. Issues have multiple attributes such as title, description, status, resolution, priority, assignee, release version, or links to other issues. Additionally, stakeholders can comment on issues and track their life-cycle. Depending on the issue tracking system and the users, there can be a variety of different link types. Common types are parent-child, duplicate, dependency, similarity, and work breakdown. In this work, we refer to dependencies between issues as \textit{links} adhering to the terminology in Jira. 

Keeping all information regarding issues visible for the user is hard in large issue tracking systems. Especially in open-source projects where a community submits hundreds of tickets weekly~\cite{stanik2018simple}. Organizing new issues and planning releases becomes a demanding, time-consuming task: a frustrating situation for developers, project managers, and product owners, as described by Fucci et al.~\cite{Fucci2018}. 

The Qt Company develops a cross-platform application framework under both commercial and open source licenses. It uses a Jira instance that everybody can submit tickets to as long as an account is created beforehand. Qt's Jira contains bugs, as well as requirements. Qt's public Jira contains over 111,959 issues, out of which 27,462 have at least one outgoing link, totaling 24,857 links in the system as of May 2019. 

In Jira's view issue pages, users only see a list of all direct links of a selected issue and not further beyond. For example, in the view issue page of QTBUG-55604 in Qt's Jira, only three issues are visible to users. However, there are 19 issues linked transitively to the issue. Such a map of links can become complex. For instance, there is a link map of size 6755 in which the longest shortest distance is 51 links across multiple Jira projects. Abad et al. presented a systematic literature review about visualization in requirements engineering~\cite{Abad2016}. The described problem can be elevated to an extent with visualization techniques. Due to the size of Qt's Jira, this alone is not sufficient to solve the problem as there are large networks. Strategies to deal with this information overload need to be developed.

Links can be unknown. A simple example is that a bug might get reported by two different users and the issues are not linked as duplicates. As links may pose constraints to release plans. It is thus important for users to be aware of relevant links as they might point to inconsistencies in future releases. Recommender systems can facilitate this task by looking for comments of users pointing out a link, using NLP to predict links between a pair of issues, or checking the consistency of an issue map concerning the release plan.

Existing Jira plug-ins with visualization for links in Jira do not offer detection for dependencies or checking consistency for releases. Combining these features with visualization such that users can make informed choices is novel in our tool.

%

\begin{figure}
\centering
\includegraphics[width=\columnwidth]{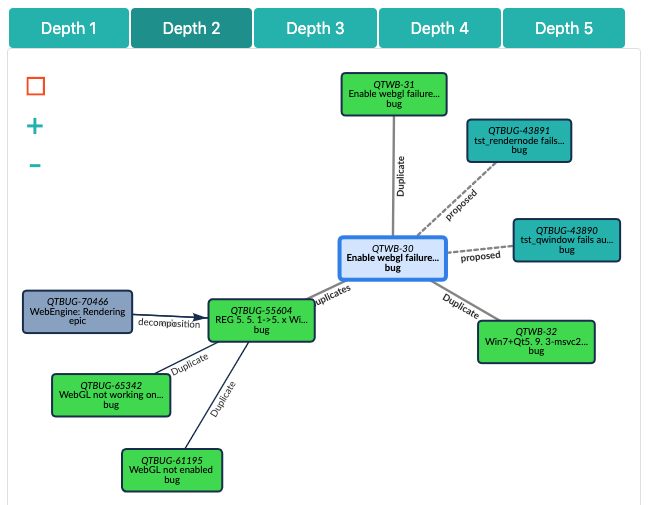}
\caption{Issue Map with depth 2 of QTBUG-30 in OpenReq Issue Link Map and Link Detection active}
\label{fig:IssueMap}
\end{figure}

With this tool, we aim to ease the described problem by combining visualization with recommender systems. The visualization (see Figure ~\ref{fig:IssueMap}) helps understand existing issues and links. It also supports decision making when rejecting or accepting a recommended link from a recommender system. The tool is part of the European Horizon 2020 project OpenReq (https://openreq.eu/). One of the project's goal is to develop an open-source Jira plug-in that any organization can use to help manage its projects. We are evaluating the current implementation of the tool with The Qt Company in their Jira instance.

\section{Overall Architecture}
The OpenReq Issue Link Map is currently a service-based tool that visualizes the link map of issues in Qt's Jira. We intend to have it as a Jira plug-in at the end of the project. So far, to get a better understanding of the whole picture, users have to explore the links one after another in Jira. With the Issue Link Map visualization users can see all linked issues at a glance. The front-end is a web-based interface while the back-end consists of several OpenReq microservices that work together in a choreographic manner. 

\section{Background Services}

\subsection{Graph of Links}
Based on issues and their links, a graph of links facilitates the visualization of an issue link map. The graph is maintained as a separate process to allow for a real-time response of a link map. This is necessary for a synchronous issue retrieval and analysis in Jira directly, since many Jira instances contain a large number of issues. 

\subsection{Link Detection}\label{dd}
Additional links are detected from the issue. To find duplicates, all issues of a single project are taken and their title and description are compared against each other to decide if they have similar content. 
Moreover, a cross-reference detection checks the comments of an issue for the mention of another issue, which can indicate a link. This happens in open-source settings: Everybody can create issues, but not everybody can create links between issues. To work around this, users often comment if they think a link exists. A maintainer then has to add the link manually. 

\subsection{Consistency Checker}\label{cc}
The consistency checker~\cite{DBLP:conf/splc/RaatikainenTMFS18} verifies that the release plan of an issue link map is consistent by applying constraint solving technologies:
First, all child issues, which have the same or higher priority, must not be assigned to a later release. Second, any required issue must not have a later release or lower priority. Third, all links from a duplicated issue are inherited by the duplicate issue.

\section{User Interface}
The user interface consists of two main parts, the visualization and the functionalities explained in~\ref{dd} and~\ref{cc}. As seen in Figure~\ref{fig:IssueMap} the link map is shown as a graph on the left side where a user can select the depth of the map. \textit{Depth} means the distance from the selected issue. The link map can be navigated by clicking any issue on the map. 
On the right-hand side, users can see general information about the issue from Jira. 
Above the information box, users can switch to the two features: link detection and consistency checker. When users call the link detection, the system presents a list of five recommended links that can be accepted or rejected. If they accept the link, they must select its type as well. 
When the users click the consistency checker they see if the issues contained in the issue link map and the corresponding releases are consistent. The different releases are also shown under the result.
Additionally, users are able to filter the issues in the link map by different attributes.

%




\section{Future Work}
The tool is still under development and thus the capabilities of the link detection can be improved. For instance, the choice of the users to accept and reject links can be used to train and refine the system, as a human-in-the-loop approach. Additionally, it is planned to visualize what part makes a certain issue link map inconsistent with the option to automatically repair the inconsistency.
The Qt trial started in January of 2018 and the tool is being developed and evaluated in this trial.
We use AI technologies in this trial to enhance issue tracking systems and make them more manageable. In the long run, we plan to also analyze user acceptance of AI technologies. 


\section*{Acknowledgement}
The work presented in this paper has been conducted within the scope of the Horizon 2020 project OpenReq, which is supported by the European Union under the Grant Nr. 732463.

\bibliographystyle{abbrv}
\bibliography{lib}
\newpage

\end{document}